\documentclass[11pt]{article}

\usepackage{amsmath,amssymb,amsthm,amsfonts,natbib,xcolor,float,thmtools,dsfont}
\usepackage[left=2.6cm,top=2.9cm,right=2.6cm,bottom=2.6cm,head=3cm,headsep=1.3cm, foot=1.8cm]{geometry}

\usepackage[bottom,flushmargin,hang,multiple]{footmisc}
\usepackage{xr-hyper}

\definecolor{darkblue}{rgb}{0.1,0.1,0.7}
\definecolor{darkred}{rgb}{0.9,0.1,0.1}

\usepackage{hyperref}
\hypersetup{colorlinks, allcolors= darkblue}

\title{On the Continuity of the\vspace{0.2cm}\\Feasible Set Mapping in Optimal Transport\vspace{0.2cm}}
\newtheorem{theorem}{Theorem}[section]
\newtheorem{proposition}[theorem]{Proposition}
\newtheorem{lemma}[theorem]{Lemma} 

\newtheorem{remark}[theorem]{Remark}

\numberwithin{equation}{section}

\newcommand{\ve}{\varepsilon}
\newcommand{\cX}{\mathcal{X}}
\newcommand{\cL}{\mathcal{L}}
\newcommand{\cF}{\mathcal{F}}
\newcommand{\cY}{\mathcal{Y}}

\newcommand{\cP}{\mathcal{P}}

\newcommand{\cS}{\mathcal{S}}
\newcommand{\cB}{\mathcal{B}}

\renewcommand{\(}{\left(}
\renewcommand{\)}{\right)}

\newcommand{\1}{\mathds{1}}

\defcitealias{BaselMarketRiskCharge}{BCBS, [2019]}

\author{Mario Ghossoub\thanks{Department of Statistics and Actuarial Science, University of Waterloo,
\href{mailto:mario.ghossoub@uwaterloo.ca}{mario.ghossoub@uwaterloo.ca}.
}
\and David Saunders\thanks{Corresponding author. Department of Statistics and Actuarial Science, University of Waterloo, \ \href{mailto:dsaunders@uwaterloo.ca}{dsaunders@uwaterloo.ca}.\vspace{0.2cm}}}

\begin{document}
\maketitle


\begin{abstract}
Consider the set of probability measures with given marginal distributions on the product of two complete, separable metric spaces, seen as a correspondence when the marginal distributions vary. In problems of optimal transport, continuity of this correspondence from marginal to joint distributions is often desired, in light of Berge's Maximum Theorem, to establish continuity of the value function in the marginal distributions, as well as stability of the set of optimal transport plans. \citet{Bergin} established the continuity of this correspondence, and in this note, we present a novel and considerably shorter proof of this important result. We then examine an application to an assignment game (transferable utility matching problem) with unknown type distributions.
\end{abstract}

\vspace{0.4cm}

{\small 

\vspace{0.4cm}

}

\vspace{0.2cm}


\section{Introduction}
Optimization problems over sets of probability measures with given marginals, and optimal transport problems in particular, arise in several contexts in economics (see, e.g., \citet{Galichon} for a book-length treatment, and the two special issues in volumes 42(2) and 67(2) of \textit{Economic Theory}). Such ubiquitous problems can be formulated as 
\begin{equation}\label{GeneralOTprimal}
\underset{\pi\in\Pi_{\cX,\cY}\(\mu,\nu\)} \sup \ \int_{\cX \times \cY} \Phi\(x,y\)\, d\pi\(x,y\),
\end{equation} 
where $\Pi_{\cX,\cY}\(\mu,\nu\)$ denotes the set of all probability measures on a product space $\cX \times \cY$ with given marginal distributions $\mu$ on $\cX$ and $\nu$ on $\cY$ (called the set of couplings of $\mu$ and $\nu$), 
and $\Phi: \cX \times \cY \rightarrow \mathbb{R}$ is a given function. 

\vspace{0.4cm}

Hereafter, $\cX$ and $\cY$ are two Polish (i.e., complete, separable, metric) spaces, with respective Borel $\sigma$-algebras $\mathcal{B}_{\cX}$ and $\mathcal{B}_{\cY}$. For a Polish space $\cS$, 
$\cP(\cS)$ is the set of all Borel probability measures on $\cS$. Given $\mu\in\cP(\cX)$, $\nu\in\cP(\cY)$, it follows that
\begin{equation*}
\Pi_{\cX,\cY}\(\mu,\nu\) = \Big\{\pi \in \cP(\cX \times \cY):  \; \pi\(A \times Y\) = \mu\(A\), \ \pi\(X \times B\) = \nu\(B\), \ \forall \(A,B\) \in \mathcal{B}_{\cX} \times \mathcal{B}_{\cY} \Big\}.
\end{equation*}

\vspace{0.4cm}

For sequences $\{\pi_{n}\}_n\subset\cP(\cS)$, $\pi_{n}\to\pi$ denotes convergence in the narrow topology on $\cP(\cS)$ (i.e., $\int f \,d\pi_{n}\to \int f \,d\pi$ for all $f\in C_{b}(\cS)$, the space of bounded continuous functions from $\cS$ to $\mathbb{R}$), which we note is metrizable by the Prokhorov metric (e.g., \citet[Theorem 6.8]{BillingsleyCPM}) on $\cP(\cS)$ defined by
\begin{equation}\label{EqProkMetric}
d_{\cP}(P,Q) :=  \inf\Big\{ \ve > 0: \; P\(A\) \leq Q\(A^{\ve}\) + \ve, \ \forall A \in \mathcal{B}_{\cS} \Big\},
\end{equation}
where $ \mathcal{B}_{\cS}$ denotes the Borel $\sigma$-algebra on $\cS$, and for each $A \in \mathcal{B}_{\cS}$, 
$$A^{\ve} := \Big\{y \in \cS: d_{\cS}\(x,y\) < \ve, \ \hbox{for some } x \in A\Big\}.$$

\noindent Then for each $\(\mu,\nu\) \in \cP\(\cX\) \times \cP\(\cY\)$, $\Pi_{\cX,\cY}\(\mu,\nu\)$ is nonempty, convex, and compact in the narrow topology on $\cP(\cX \times \cY)$ (e.g., \citet[pp.\ 32, 49-50]{VillaniTopicsInOT}).
 
\vspace{0.4cm} 

Problem \eqref{GeneralOTprimal} is precisely the Monge-Kantorovich optimal transport problem. Here, we are interested in the properties of the correspondence $\Pi_{\cX,\cY} : \cP(\cX)\times \cP(\cY) \twoheadrightarrow \cP(\cX\times \cY)$. Formally,  $\Pi_{\cX,\cY}$ associates to each pair $\(\mu,\nu\) \in \cP(\cX)\times \cP(\cY)$ of marginal distributions the feasibility set $\Pi_{\cX,\cY}\(\mu,\nu\)$ of Problem \eqref{GeneralOTprimal}. Let $Gr\(\Pi_{\cX,\cY}\)$ denote the graph of the correspondence $\Pi_{\cX,\cY}$, given by
\begin{equation*}
Gr\(\Pi_{\cX,\cY}\) :=  \Big\{ \(\(\mu,\nu\),\pi\) \in \(\cP(\cX)\times \cP(\cY)\) \times \cP(\cX\times \cY): \; \pi \in \Pi_{\cX,\cY}\(\mu,\nu\)\Big\}.
\end{equation*}
We define the linear functional $\Psi: Gr\(\Pi_{\cX,\cY}\) \rightarrow \mathbb{R}$ by 
\begin{equation}\label{ObjectiveFct}
\Psi\(\(\mu,\nu\),\pi\) := \int_{\cX \times \cY} \Phi\(x,y\)\, d\pi\(x,y\). 
\end{equation}
Furthermore, we define the value function $V: \cX\times \cY \rightarrow \mathbb{R}$ for Problem \eqref{GeneralOTprimal} by
\begin{equation}\label{ValueFct}
V\(\mu,\nu\) := \underset{\pi \in \Pi_{\cX,\cY}\(\mu,\nu\)} \sup \Psi\(\(\mu,\nu\),\pi\).
\end{equation}

\vspace{0.1cm}

\noindent Finally, we define the correspondence $\mathcal{M}: \cP(\cX)\times \cP(\cY) \twoheadrightarrow \cP(\cX\times \cY)$, which assigns to each given pair of marginal distributions $\(\mu,\nu\)$ the set of optimizers of Problem \eqref{GeneralOTprimal}, by
\begin{equation}\label{argmaxCorr}
\mathcal{M} := \Big\{ \pi^* \in \Pi_{\cX,\cY}\(\mu,\nu\): \; \Psi\(\(\mu,\nu\),\pi^*\) = V\(\mu,\nu\)\Big\}.
\end{equation}

\noindent Note that, by the Monge-Katorovich Duality Theorem (e.g., \citet[Theorem 1.3]{VillaniTopicsInOT}), nonemptiness of $\mathcal{M}$ follows from the upper-semicontinuity of the function $\Phi$, as long as there are lower-semicontinuous functions $a \in L^1\(\cX,\mathcal{B}_{\cX}, \mu\)$ and $b \in L^1\(\cY,\mathcal{B}_{\cY}, \nu\)$ such that $\Phi\(x,y\) \leq a\(x\) + b\(y\)$, for $\mu$-a.e.\ $x$ and $\nu$-a.e.\ $y$.

\vspace{0.4cm}

One is typically interested in properties of the correspondences $\Pi_{\cX,\cY}$ and $\mathcal{M}$, as well as continuity of the value function $V$, which is important when approximating Problem \eqref{GeneralOTprimal} in practice. Moreover, while it is immediate to see that $\Pi_{\cX,\cY}$ has nonempty, convex, and compact values in the narrow topology on $\cP(\cX\times \cY)$, the continuity of $\Pi_{\cX,\cY}$ is of primary concern, in light of Berge's Maximum Theorem. 
Indeed, since $\Pi_{\cX,\cY}$ has nonempty compact values, and since $\cP(\cX\times \cY)$ is Hausdorff, being metrizable, continuity of the value function of Problem \eqref{GeneralOTprimal} and upper hemicontinuity of the correspondence $\mathcal{M}$ would follow from continuity of the correspondence $\Pi_{\cX,\cY}$, under mild regularity conditions on the function $\Phi$.

\vspace{0.4cm}

\citet{Bergin} and \cite{SavchenkoZarichnyi2014} provided proofs of the continuity of the feasible set correspondence $\Pi_{\cX,\cY}$ based on rather lengthy arguments. In this paper, we present in Section \ref{sectionMainResult} an alternative, much shorter proof of this important result, using well-known measure theoretic tools. We then examine in Section \ref{sectionApplication} an application to a canonical matching problem with transferable utility.

\vspace{0.4cm}


\section{Continuity of the Feasible Set Correspondence $\Pi_{\cX,\cY}$}
\label{sectionMainResult}

We will make use of the following two results. The first can be found in \citet[Theorem 3.1.2]{EthierKurtz}), and it provides a useful alternative characterization of the metrizability of narrow convergence. The second can be found in \citet[pp.\ 208-210]{VillaniTopicsInOT}), and it is often referred to as the Gluing Lemma.
 
\vspace{0.2cm}

\begin{lemma}\label{LemProkMetric}
Let $\(\cS, d_{\cS}\)$ be a Polish space with Borel $\sigma$-algebra $\mathcal{B}_{\cS}$ and let $d_{\cP}$ denote the Prokhorov metric on $\cP(\cS)$, defined in eq.\ \eqref{EqProkMetric}. Then
\begin{equation}\label{EqProkMetric2}
d_{\cP}(P,Q) = \inf_{m\in\Pi_{\cS,\cS}\(\mu,\nu\)} \inf\Big\{ \ve > 0: \; m[(x,y): d_{\cS}(x,y) \geq \ve ] \leq \ve\Big\}.
\end{equation} 
\end{lemma}

\vspace{0.2cm}

\begin{lemma}[Gluing Lemma]\label{GluingLemma}
Let $v_{1}$, $v_{2}$, $v_{3}$ be three probability measures supported in Polish spaces $\cS_{1}$, $\cS_{2}$, $\cS_{3}$ respectively, and let $m_{12} \in \Pi_{\cS_{1},\cS_{2}}\(v_{1},v_{2}\)$ and $m_{23}\in \Pi_{\cS_{2},\cS_{3}}\(v_{2},v_{3}\)$ be two transference plans. Then there exists a probability measure $m\in\cP(\cS_{1}\times \cS_{2}\times \cS_{3})$ with marginals $m_{12}$ on $\cS_{1}\times \cS_{2}$ and $m_{23}$ on $\cS_{2}\times \cS_{3}$. That is, if $\mathcal{B}_{\cS_{1} \times \cS_{2}}$ and $\mathcal{B}_{\cS_{2} \times \cS_{3}}$ denote the Borel $\sigma$-algebras of $\cS_{1} \times \cS_{2}$ and $\cS_{2} \times \cS_{3}$, respectively, then
\begin{equation*}
m\(A \times \cS_3\) = m_{12}\(A\), \ m\(\cS_1 \times B\) = m_{23}\(B\), \ \forall \(A,B\) \in \mathcal{B}_{\cS_{1} \times \cS_{2}} \times \mathcal{B}_{\cS_{2} \times \cS_{3}}.
\end{equation*}
\end{lemma}

\vspace{0.2cm}

\begin{theorem}\label{OTCorrespondenceContinuity}  
The correspondence $\Pi_{\cX,\cY} : \cP(\cX)\times \cP(\cY) \twoheadrightarrow \cP(\cX\times \cY)$ is continuous, and has nonempty, convex, and compact values in the narrow topology on $\cP(\cX\times \cY)$.
\end{theorem}

\vspace{0.1cm}

\begin{proof}  
First, note that for every $\(\mu,\nu\) \in\cP(\cX) \times \cP(\cY)$, $\Pi_{\cX,\cY}\(\mu,\nu\) \neq \varnothing$, since the tensor product $\mu \otimes \nu$ belongs to  $\Pi_{\cX,\cY}\(\mu,\nu\)$. Moreover, $\Pi_{\cX,\cY}$ trivially has convex values. Compactness of the values of $\Pi_{\cX,\cY}$ in the narrow topology on $\cP(\cX\times \cY)$ is shown in \citet[pp.\ 49-50]{VillaniTopicsInOT}, for instance. We now show continuity of $\Pi_{\cX,\cY}$.

\vspace{0.2cm}

To show upper hemicontinuity, suppose that we have $\{\(\(\mu_n,\nu_n\),\pi_n\)\}_n \subset Gr\(\Pi_{\cX,\cY}\)$, with $\mu_{n}\to\mu$ and $\nu_{n}\to\nu$. Hence, $\{\mu_{n}\}_n$ and $\{\nu_{n}\}_n$ and tight, by Prokhorov's Theorem (e.g., \citet[Theorems 5.1-5.2]{BillingsleyCPM}). Tightness of $\{\mu_{n}\}_n$ and $\{\nu_{n}\}_n$ implies that of $\{\pi_{n}\}_n$, so that by Prokhorov's Theorem there exists a convergent subsequence $\pi_{n_{k}}\to\pi$. For any $\(f,g\) \in C_{b}(\cX) \times  C_{b}(\cY)$, we have 
\begin{equation*}
\begin{split}
&\int f\, d\pi = \lim_{k\to\infty} \int f \, d\pi_{n_{k}} = \lim_{k\to\infty} \int f\, d\mu_{n_{k}} = \int f\, d\mu;\\
&\int g\, d\pi = \lim_{k\to\infty} \int g \, d\pi_{n_{k}} = \lim_{k\to\infty} \int g\, d\nu_{n_{k}} = \int g \, d\nu.
\end{split}
\end{equation*}

\noindent Therefore, 
$$\int_{\cX \times \cY} \left[f + g\right] \, d\pi = \int_\cX f\, d\mu +  \int_\cY g\, d\nu,$$
and since $\cX, \cY$ are Polish spaces, it follows from \citet[p.\ 18]{VillaniTopicsInOT} that $\pi\in\Pi_{\cX,\cY}\(\mu,\nu\)$.

\vspace{0.4cm} 
 
To show lower hemicontinuity, fix $\pi\in\Pi_{\cX,\cY}\(\mu,\nu\)$ and suppose that we have $\{\(\mu_n,\nu_n\)\}_n \subset \cP(\cX)\times \cP(\cY)$, with $\mu_{n}\to\mu$ and $\nu_{n}\to\nu$. Since $\mu_{n} \to \mu$, it follows that $d_{\cP}(\mu_{n},\mu) \to 0$, where $d_{\cP}$ denotes the Prokhorov metric, characterized in Lemma \ref{LemProkMetric}. Fix $n$,  let $0< \ve_{n} \leq d_{\cP}(\mu_{n},\mu) + \tfrac{1}{n}$, and let $v_{1,n}\in\Pi_{\cX,\cX}(\mu_{n},\mu)$ be such that $v_{1,n}\left[(x,x'): d_{\cX}(x,x')\geq\ve_{n}\right] \leq \ve_{n}$. Applying Lemma \ref{GluingLemma} with $\cS_{i}=\cX$ for $i=1,2$, $\cS_{3}=\cY$, $\pi_{12} = v_{1,n}$, and $\pi_{23} = \pi$, we obtain a measure $m_{1,n}$ on $\cX \times \cX \times \cY$ with the required ``bivariate" marginal distributions. 

\vspace{0.2cm}

Similarly, let $0< \delta_{n} \leq d_{\cP}(\nu,\nu_{n}) + \tfrac{1}{n}$ and $v_{2,n}\in \Pi_{\cY,\cY}\(\nu,\nu_{n}\)$ be such that $v_{2,n}[(y,y'): d_{\cY}(y,y')\geq\delta_{n}] \leq \delta_{n}$. Apply Lemma \ref{GluingLemma} again with $\cS_{1} = \cX\times\cX$, $\cS_{2}=\cS_{3}=\cY$, $\pi_{12} = m_{1,n}$, and $\pi_{23} = v_{2,n}$ to obtain a measure $m_{n}$ on $\cX \times \cX \times \cY \times \cY$ with ``univariate" marginal distributions $\mu_{n},\mu,\nu,\nu_{n}$ and ``bivariate" marginal distributions $v_{1,n}$ for the first and second components, $\pi$ for the second and third components, and $v_{2,n}$ for the third and fourth components. 

\vspace{0.2cm}

Let $\pi_{n}$ denote the ``bivariate" marginal distribution (on $\cX\times\cY$) of the first and fourth components (so that $\pi_{n} \in \Pi_{\cX,\cY}\(\mu_{n},\nu_{n}\)$), and consider the measure $\widetilde m_{n}$ on $(\cX \times\cY) \times (\cX \times \cY)$ with marginals $(\pi_{n},\pi)$ that is the image of $m_{n}$ under the 
mapping $\sigma(x_{1},x_{2},y_{1},y_{2}) = (x_{1},y_{2},x_{2},y_{1})$. Metrize the product space using $$d_{\cX \times \cY}((x,y),(x',y')) := \max(d_{\cX}(x,x'),d_{\cY}(y,y')),$$ so that $\Big\{d_{\cX \times \cY}((x,y),(x',y')) \geq c\Big\} \Longrightarrow \Big\{d_{\cX}(x,x') \geq c$ or $d_{\cY}(y,y') \geq c$\Big\}. Then 
\begin{equation*}
\begin{split} 
\widetilde m[ d_{\cX \times \cY}((x,y),(x',y')) \geq \ve_{n} + \delta_{n}] 
& \leq \widetilde m [ d_{\cX}(x,x') \geq \ve_{n} + \delta_{n} ] + \widetilde m [ d_{\cY}(y,y') \geq \ve_{n} + \delta_{n}] \\
& \leq \widetilde m[ d_{\cX}(x,x') \geq \ve_{n}] + \widetilde m [ d_{\cY}(y,y') \geq \delta_{n}] \\
& \leq \ve_{n} + \delta_{n}.
\end{split} 
\end{equation*}
Thus $d_{\cP}(\pi_{n},\pi) \leq \ve_{n} + \delta_{n}$ , $\pi_{n} \in \Pi_{\cX,\cY}\(\mu_{n},\nu_{n}\)$, and $\pi_{n} \to \pi$.
\end{proof}

\vspace{0.4cm}



%
%
%
%
%
%
%
%
%
%
%
%

\section{TU Matching with Unknown Type Distributions}
\label{sectionApplication}

We consider a canonical example of a matching problem with transferable utility, or assignment game (\cite{ShapleyShubik1971}), in which a central planner seeks to assign an element $x$ from a population $\cX$ to an element $y$ from a population $\cY$. Both $\cX$ and $\cY$ can be multidimensional, and we take them to be generic nonempty Polish spaces. The spaces $\cX$ and $\cY$ are equipped with Borel probability measures $\mu$ and $\nu$, respectively, representing the distribution of agents' types over the respective spaces.  Let $\Phi: \cX \times \cY \rightarrow \mathbb{R}$ denote the joint utility (or surplus) function, whereby $\Phi\(x,y\)$ is the joint surplus generated if $x \in \cX$ is matched with $y \in \cY$. For instance, $\mu$ can denote the distribution of skills over a set $\cX$ for a population of workers, $\nu$ the distribution of firm characteristics over a set $\cY$, and $\Phi\(x,y\)$ denotes the value created if a worker with skill $x \in \cX$ is employed by a firm with characteristic $y \in \cY$. 

\vspace{0.2cm}

Following \cite{Chiapporietal2010}, an \textit{assignment} of $x \in \cX$ to $y \in \cY$ is a probability measure $\pi\in\Pi_{\cX,\cY}\(\mu,\nu\)$ with support $supp\(\pi\) \subset \cX \times \cY$, which leads to an economic value, or total surplus of $$\int_{\cX \times \cY} \Phi\(x,y\)\, d\pi\(x,y\).$$ 

\noindent A \textit{payoff} corresponding to an assignment $\pi\in\Pi_{\cX,\cY}\(\mu,\nu\)$ is a pair of functions
$\(U_{\cX},U_{\cY}\) \in L^1\(\cX,\mathcal{B}_{\cX}, \mu\)\times L^1\(\cY,\mathcal{B}_{\cY}, \nu\)$ such that
$$U_{\cX}(x) + U_{\cY}(y) =  \Phi\(x,y\), \ \hbox{for $\pi$-a.e.\,} \(x,y\) \in supp\(\pi\).$$

\noindent An \textit{outcome} is a triple $\(\pi,U_{\cX},U_{\cY}\)$, where $\(U_{\cX},U_{\cY}\)$ is a payoff corresponding to $\pi$. The standard equilibrium concept used in this framework is \textit{satibility}. An outcome $\(\pi,U_{\cX},U_{\cY}\)$ is called \textit{stable} if it satisfies 
$$U_{\cX}(x) + U_{\cY}(y) \geq  \Phi\(x,y\), \ \forall \(x,y\) \in \cX \times \cY.$$

\noindent Finally, a matching $\pi$ is \textit{stable} if there exists a payoff $\(U_{\cX},U_{\cY}\)$ corresponding to $\pi$, such that the outcome $\(\pi,U_{\cX},U_{\cY}\)$ is stable. Hence, stability is tantamount to robustness against deviations by both individuals and pairs. In other words stability requires that (i) no matched agent is better off unmatched; and (ii) no two unmatched agents are better off matched together than remaining in their current situation. 

\vspace{0.2cm}

A fundamental result in the theory of matching with transferable utility is that stability is equivalent to surplus maximization. This result is due to \cite{ShapleyShubik1971} in the discrete case and \cite{Gretskyetal1992} in the continuous case (and it is also a consequence of the Monge-Kantorovich duality \citet[Theorem 5.10]{VillaniOTOldAndNew}). It was recently extended by \cite{Pass2019} to a setting of tripartite matching (also known as multi-marginal optimal transport).

\vspace{0.2cm}

\begin{proposition}\label{PropStab}
For a given surplus function $\Phi: \cX \times \cY \rightarrow \mathbb{R}$, a matching $\pi\in\Pi_{\cX,\cY}\(\mu,\nu\)$ is stable if and only if it solves the surplus maximization problem \eqref{GeneralOTprimal}:
\begin{equation*}\label{GeneralOTprimalEx}
\underset{\pi\in\Pi_{\cX,\cY}\(\mu,\nu\)} \sup \ \int_{\cX \times \cY} \Phi\(x,y\)\, d\pi\(x,y\).
\end{equation*}
\end{proposition}

\vspace{0.2cm}

A central planner can hence implement a stable, that is, equilibrium assignment by solving the surplus maximization problem \eqref{GeneralOTprimal}. This, however, necessitates knowledge of the marginal (type) distributions $\mu$ and $\nu$. If the type distributions $\mu$ and $\nu$ are unknown by the central planner, then since $\cX$ and $\cY$ are separable, an approximation based on sampling from empirical distributions can be used, as long as the value of Problem \eqref{GeneralOTprimal} is continuous. This, in turn, can be obtained from Berge's Maximum Theorem when the correspondence $\Pi_{\cX,\cY}$ is continuous, under some regularity conditions on the surplus function $\Phi$. We summarize this in Proposition \ref{StableApprox} below. First, however, we introduce some needed notation.

\vspace{0.2cm}

For the probability space $\(\cX, \cB_\cX, \mu\)$, there are $\cX$-valued independent random variables $\{X_i\}_{i \geq 1}$ defined on a common probability space $\(\Omega_\cX, \cF_\cX, P_\cX\)$, with laws $\cL\(X_i\) = P_\cX \circ X_i^{-1} = \mu$, for all $i \geq 1$ (see, e.g., \citet[\S 8.2, \S 11.4]{DudleyRAP}). 
Similarly, for the probability space $\(\cY, \cB_\cY, \nu\)$, there are $\cY$-valued independent random variables $\{Y_j\}_{j \geq 1}$ defined on a common probability space $\(\Omega_\cY, \cF_\cY, P_\cY\)$, with laws $\cL\(Y_j\) = P_\cY \circ Y_j^{-1} = \nu$, for all $j \geq 1$. Define the \textit{empirical measures} by 
\begin{equation}\label{EmpMu}
\mu_n(A)(\omega) := \frac{1}{n} \sum_{i=1}^n  \1_A\(X_i(\omega)\) , \ \forall A \in \cB_\cX, \ \forall \omega \in \Omega_\cX;
\end{equation}
and 
\begin{equation}\label{EmpNu}
\nu_n(B)(\kappa) := \frac{1}{n} \sum_{j=1}^n  \1_B\(Y_j(\kappa)\), \ \forall B \in \cB_\cY, \ \forall \kappa \in \Omega_\cY.
\end{equation}

\vspace{0.2cm}

\begin{proposition}\label{StableApprox}
Let $V: \cX\times \cY \rightarrow \mathbb{R}$ be the value function of Problem \eqref{GeneralOTprimal} defined in eq.\ \eqref{ValueFct}, and let $\{\mu_n\}_n$ and $\{\nu_n\}_n$ be the empricical measures defined in eq.\ \eqref{EmpMu} and \eqref{EmpNu}. If $\Phi \in C_{b}\(\cX \times \cY\)$, then there exists a stable matching $\pi^*$. Moreover, $$V\(\mu_n,\nu_n\) \to V\(\mu,\nu\) = \int_{\cX \times \cY} \Phi\(x,y\)\, d\pi^*\(x,y\) = \underset{\pi \in \Pi_{\cX,\cY}\(\mu,\nu\)} \sup \Psi\(\(\mu,\nu\),\pi\),$$
where $\Psi$  denotes the objective function of Problem \ref{GeneralOTprimal} defined in eq.\ \eqref{ObjectiveFct}. 
\end{proposition}

\vspace{0.1cm}

\begin{proof} 
First, note that by the Monge-Katorovich Duality Theorem (e.g., \citet[Theorem 1.3]{VillaniTopicsInOT}), the assumption that $\Phi \in C_{b}\(\cX \times \cY\)$ guarantees the existence of a solution $\pi^*$ to Problem \eqref{GeneralOTprimal}. Hence, by Proposition \ref{PropStab}, $\pi^*$ is a stable matching. By Theorem \ref{OTCorrespondenceContinuity}, $\Pi_{\cX,\cY}$ is continuous, implying that $Gr\(\Pi_{\cX,\cY}\)$ is closed. Since $\Phi \in C_{b}\(\cX \times \cY\)$, it follows that the objective function $\Psi$ of Problem \ref{GeneralOTprimal} is continuous in the product topology. Since $\Pi_{\cX,\cY}$ is continuous and has nonempty, compact values, and since $\cP(\cX\times \cY)$ is Hausdorff being metrizable, continuity of the value function $V$ of Problem \eqref{GeneralOTprimal} follows from Berge's Maximum Theorem (e.g., \citet[Theorem 17.31]{AliprantisBorder}). 
By Varadarajan's extension (\citet[Theorem 11.4.1]{DudleyRAP}) of the classical Glivenko-Cantelli Theorem, the sequences $\{\mu_n\}_n$ and $\{\nu_n\}_n$ converge almost surely to $\mu$ and $\nu$, respectively, since the spaces $\cX$ and $\cY$ are separable. Therefore, $\mu_n \to \mu$ and $\nu_n \to \nu$. Hence, by continuity of $V$, it follows that $V\(\mu_n,\nu_n\) \to V\(\mu,\nu\).$
\end{proof}

\vspace{0.2cm}

\begin{remark}
In light of the Monge-Katorovich Duality Theorem, 
the assumption in Proposition \ref{StableApprox}  that $\Phi \in C_{b}\(\cX \times \cY\)$ can be weakened to an assumption that $\Phi$ is upper-semicontinuous and that there are some lower-semicontinuous functions $a \in L^1\(\cX,\mathcal{B}_{\cX}, \mu\)$ and $b \in L^1\(\cY,\mathcal{B}_{\cY}, \nu\)$ such that 
\begin{equation*}\label{CondDom}
\Phi\(x,y\) \leq a\(x\) + b\(y\), \hbox{ for $\mu$-a.e.\ $x$ and $\nu$-a.e.\ $y$}.
\end{equation*}
\end{remark}

\vspace{0.2cm}

\begin{remark}[Hedonic Price Equilibria]
\cite{Chiapporietal2010} show that there exists a canonical correspondence between models of hedonic pricing with quasi-linear preferences and TU matching models, and hence \textit{a fortiori} surplus maximization problems (in light of Proposition \ref{PropStab}). This was extended by \cite{Pass2019} to a setting of multi-marginal optimal transport (tripartite matching). We refer to \cite{Ekeland2005}, \cite{Ekeland2010b}, \cite{Chiapporietal2010}, and \cite{Pass2019} for more about models of hedonic equilibria and their equivalence to surplus maximization problems. Proposition \ref{StableApprox} above can therefore be used to show the existence of a hedonic price equilibrium, when the type distributions of buyers and sellers (the probability measures $\mu$ and $\nu$) are unknown.
\end{remark}

%

%
%
%

\vspace{0.5cm}


\bibliography{References}

\end{document}